\newcommand{\msun}{\mbox{$M_{\odot}$~}}
\newcommand{\Msun}{\mbox{$M_{\odot}$~}}
\newcommand{\Mgmc}{\mbox{$M_{\rm GMC}$}}
\newcommand{\Mymc}{\mbox{$M_{\rm YMC}$}}
\newcommand{\Rgmc}{\mbox{$R_{\rm GMC}$}}
\newcommand{\Rymc}{\mbox{$R_{\rm YMC}$}}
\newcommand{\halpha}{\mbox{H$\alpha$~}}

\documentclass{aa}
\usepackage{graphicx}

\begin{document}
   \title{Hierarchical Star Formation in M51: Star/Cluster Complexes}

   \titlerunning{Star/Cluster Complexes in M51}

   \author{N. Bastian \inst{1}, M. Gieles \inst{1}, Yu. N. Efremov
   \inst{2} and H.J.G.L.M. Lamers\inst{1,3}}
   \authorrunning{Bastian, Gieles, Efremov, and Lamers}
   \offprints{bastian@astro.uu.nl}

   \institute{$^1$Astronomical Institute, Utrecht University, 
              Princetonplein 5, NL-3584 CC Utrecht The Netherlands \\
              \email{bastian@astro.uu.nl, gieles@astro.uu.nl} \\
              $^2$ Sternberg Astronomical Institute of Moscow State
              University, Universitetsky Prospect, 13, Moscow, 119899,
              Russia\\
	      \email{efremov@sai.msu.ru}\\ 
               $^3$SRON Laboratory for Space Research, Sorbonnelaan 2, 
             NL-3584 CA Utrecht, The Netherlands\\
	     \email{lamers@phys.uu.nl}\\}

   \date{Received XXX; accepted XXXX}

%   \today

   \abstract{We report on a study of young star cluster complexes in
   the spiral galaxy M51.  Recent studies have confirmed that star
   clusters do not form in isolation, but instead tend to form in
   larger groupings or complexes.  We use {\it HST} broad and narrow band
   images (from both {\it WFPC2} and {\it ACS}), along with {\it
   BIMA}-CO observations to study the
   properties and investigate the origin of these complexes.  We find
   that the complexes are all young ($< 10$ Myr), have sizes
   between $\sim$85 and $\sim$240 pc, and have masses between 3-30 $\times
   10^{4} M_{\odot}$.  Unlike that found for isolated young star clusters, we
   find a strong correlation between the complex mass and radius,
   namely $M\propto R^{2.33 \pm 0.19}$. This is similar to that found for giant
   molecular clouds (GMCs).   By
   comparing the mass-radius relation of GMCs in M51 to that of the
   complexes we can estimate the star formation efficiency within the
   complexes, although this value is heavily dependent on the assumed
   CO-to-H$_2$ conversion factor.  The complexes
   studied here have the same surface density distribution as
   individual young star clusters and GMCs.  If star formation within
   the complexes is proportional to the gas density at that point,
   then the shared mass-radius relation of GMCs and complexes is a
   natural consequence of their shared density profiles.  We briefly
   discuss possibilities for the lack of a mass-radius relation for
   young star clusters.  We note that many of the complexes show
   evidence of merging of star clusters in their centres, suggesting
   that larger star clusters can be produced through the build up of
   smaller clusters.
 }

\maketitle

%
%________________________________________________________________

\section{Introduction}
\label{intro}
Recent studies of star cluster populations have shown that young star
clusters do not form in isolation, but tend to be clustered
themselves (Zhang et al.~2001, Larsen 2004).  Despite the
relatively large amount of attention that young star clusters have
received in recent years, these complexes have been largely ignored,
leaving many of their basic properties unknown.  What triggers the
formation of these complexes?  What are their masses and sizes?
Do their properties resemble those of giant molecular clouds (GMCs)
from which they are formed or are they more akin to
single young massive star clusters?    It is this last
question that may shed light on their formation mechanism.

Star/cluster complexes are the largest and oldest objects in the
hierarchy of embedded groupings,
which starts with multiple stars and finishes on galactic scales. The
younger and smaller complexes are always within the older and the
larger ones. This hierarchy is similar to the fractal distribution observed
in the interstellar gas and in fact is the result of the latter
(Elmegreen and Efremov 1996; Efremov and Elmegreen 1998; Elmegreen, 2004).

Large star forming complexes associated with the spiral arms of disk
galaxies are well known.  Analytic considerations by Elmegreen (1994)
showed that strong spiral arms will trigger GMC formation within them,
which in turn may lead to the formation of star/cluster complexes. The
range in age of objects within these star/cluster complexes is usually 
quite small ($<20$ Myr) suggesting a coherent formation mechanism,
which separates them from other large star 
forming regions within disk galaxies (such as the local Gould Belt in
the Galaxy) which have a much larger intrinsic spread in age within
them (Efremov 1995).  Elmegreen \& Elmegreen (1983) suggested that
complexes associated with spiral arms form
from large HI/CO clouds due to gravitational instabilities along the
spiral arm.  If the complexes formed directly from large gas clouds we
may expect them to retain some of the properties of their progenitor
clouds.  Additionally, the high pressures associated with
spiral arms are conducive to the formation of massive star clusters
instead of loose associations (Elmegreen \& Efremov 1997).  Thus we
may expect complexes associated with spiral arms to contain a rich
population of compact stellar clusters.

Galactic giant molecular clouds (GMCs) have a clear relation between
their mass and 
radius, namely \Mgmc~$\propto$~\Rgmc$^{2}$ (Solomon et al. 1987). This
relation holds down to the scale of cloud clumps, which are only a few
parsecs in radius (Williams et al. 1995).  Young massive
star clusters (YMCs), on the other hand, show a weak relation
between cluster mass and radius, namely \Mymc~$\propto$~\Rymc$^{0.1}$,
with a large scatter (Zepf et
al. 1999; Larsen 2004;
Bastian et al. 2005a\footnote{The data of Bastian et al. 2005a, of
young star clusters in M51, are
consistent with no relation between the mass and radius.}). Assuming
that virialized GMCs are the 
progenitors of young massive clusters, this result is quite
surprising because it implies that this imprint from the parent GMC
(i.e. the mass-radius relation) must be erased on 
timescales similar to the formation timescale of the clusters.  Thus,
a mechanism which destroys any initial mass-radius 
relation must be a key ingredient in star cluster formation models
(Ashman \& Zepf 2001).  In this paper, we look at one scale larger
than the individual star clusters, namely that of clusters of
clusters, or cluster complexes.  

An important question about these complexes concerns their
future evolution.  $N$-body simulations (Kroupa 1998) have shown
that in complexes of high cluster densities, significant merging
of clusters is likely.  This may provide a mechanism for the formation
of extremely massive clusters (Fellhauer \& Kroupa 2002, Fellhauer \&
Kroupa 2005).  On the other hand,
Bastian et al.~(2005b) have shown that gas is being expelled 
from complexes in the Antennae galaxies ($\sim10-40$ km/s).
This will result in a rapid decrease in binding energy of the systems,
and will tend to destroy the complexes.  We will address this point by
studying the properties of clusters within the complexes.

This paper is organised in the following way.  In \S~2 we present both
the optical and CO observations and introduce the complexes in M51.
\S~3 is dedicated to the derivation 
of the properties of each complex (e.g. their size, age, mass and star
formation rates), as well as an analysis of
the cluster population within each complex.  In \S~4 we discuss the
formation and evolution of the cluster complexes in light of their
relation to young  massive clusters and giant molecular clouds.
Finally, in \S~5 we discuss the complexes in terms of the
general hierarchy of star formation within galaxies and in
\S~\ref{conclusions} we summarize the main results.

\section{Observations}
\label{obs}
\subsection{{\it HST WFPC2} and {\it ACS} observations}
The observations used in this study were taken from the {\it HST}
archive, and are presented in detail in Bastian et al. (2005a).  The
data set consists of two pointings, shown in Fig.~1 in Bastian et
al. (2005a), each consisting of broadband F439W ($\approx$ B), F555W
($\approx$ V), F675W ($\approx$ R), F814W ($\approx$ I), and F656N
(H$\alpha$) filters.  In addition, Field~2 also has F336W ($\approx$
U) filter
observations, which are crucial for age dating young star clusters.
Thus, we will concentrate the detailed analysis on those complexes
which have F336W observations, while using the others to
corroborate the results.

We have also used the {\it Hubble Heritage ACS} images which
cover the entire optical galaxy in F435W (B), F555W (V), F658N
(H$\alpha$), and F814W (I) filters.  For a full review of the exposure
and reduction information, see Mutchler et al. (2005).  Due to
the large coverage and 
high spatial resolution, this data will mainly be exploited to obtain
size estimates of the individual sources within the complexes.

The complexes were found using flux contour cuts on the F439W, F555W
and F675W images.  In the regions away from the centre of the galaxy,
this resulted in a fairly unambiguous selection.  Complexes in the inner
spiral regions were discarded from our sample, as background
variations made the identification of the complexes non-trivial.  The
one exception to this is a large complex in the north-eastern section
of the inner spiral arms, as this complex was noted in the study of
Scoville et al. (2001).  These authors also noted the existence of
Complex~G2 in their study.  The complexes are identified in
Fig.~\ref{mosaic1}.   The numbering system (1 or 2) corresponds to the two
different spiral arms.  Complexes with F336W observations are~B1, C2,
D2, E2, F2, and G2. 

The magnitude of each complex was measured with a circular aperture
with a size set equal to that of the complex (see \S~\ref{sizes}).
The background was determined and subtracted using a ring with inner
radius 6 pixels more than the radius of the complex, and with a width
of 10 pixels.  The magnitudes of the individual sources within the
complexes were measured using the PSF fitting package {\it HSTphot}
(Dolphin 2000).  This
is different from the technique used in Bastian et al. (2005a), who used
aperture photometry.  The change is due to the increased crowding within the
complexes relative to the field. 

The positions of the complexes are shown in Fig.~\ref{mosaic1}, while
Figs.~\ref{complexes-f1}~\&~\ref{complexes-f2} show enlarged images of
the complexes without and with F336W($\sim U$)-band data respectively.
Additionally, we show an {\it HST-ACS} colour composite image of the
largest complex in our
sample (Complex G2) along with an image of the complex containing the largest
super-star cluster (Complex A1) in Fig.~\ref{g-rgb}, where blue, green
and red represent the F439W~($\approx B$), F555W ($\approx V$) and F656N
($\approx H\alpha$) respectively.

Throughout this paper, we assume the same distance to M~51 which was
used in Bastian et al.~(2005a), namely 8.4~Mpc.

\begin{figure*}[tbh]
\begin{center}
\end{center}
\caption{Positions of the cluster complexes overlayed on
the F555W (V) band image. The complexes are identified with circles
representing their measured radii as determined in \S~\ref{sizes}.
The image is 260~$\times$~170~arcseconds$^{2}$}  
\label{mosaic1}
\end{figure*}

        \begin{figure}[tbh]	
	\begin{center}
       \caption{F555W images of the complexes without F336W data.   The images
       are 120 $\times$ 120 pixels$^{2}$ (12 by 12 arseconds) which
	corresponds to $\sim 
       550 \times 550 {\rm pc}^{2}$. The circles on the images correspond to
       the radius derived for each complex.}   
        \label{complexes-f1}
      \end{center} 
        \end{figure}        
 
         \begin{figure}[tbh]	
	 \begin{center}
        \caption{F555W images of complexes with F336W data.  The image of
       Complex C2 is 70 $\times$ 70 pixels$^{2}$ which is approximately
       300 by 300 pc$^2$.  The rest of the images are 120 $\times$ 120
       pixels$^{2}$ (12 by 12 arcseconds) which corresponds to $\sim
	 550 \times 550 
       {\rm pc}^{2}$.  The circles on the images correspond to
       the radius derived for each complex.}  
        \label{complexes-f2}
      \end{center}
               \end{figure}        
 
         \begin{figure}	
	 \begin{center}

        \caption{Composite {\it HST-ACS} images of complex G2 (top) and A1
       (bottom).  The top panel is $\sim475$~pc on a side, while the
       bottom panel is $\sim800$~pc on a side.  The circles in the
       bottom panel outline the outer and inner dust arcs, and
       shows that neither of them are centred on the young
       central star cluster (the centre of the arcs is marked by a
       cross).  The blue, green, and  
       red correspond to F439W, F555W, and F656N filters
       respectively.}  
        \label{g-rgb}
      \end{center}
               \end{figure}

\subsection{{\it Bima CO} observations}
\label{sec:bima}
We obtained CO (J=1-0) observations of M51 from the {\it BIMA SONG}
(Survey of Nearby Galaxies) survey\footnote{The data were taken from
the {\it NASA Extragalactic Database}, which can be found at
http://nedwww.ipac.caltech.edu/}.  The data are presented in detail 
in Helfer et al. (2003).  As described in Henry et al. (2003), we used
an image of M51 which was obtained by integrating the intensity over all
velocity channels.  The details of the observations and reduction of
the data are given in detail in the preceding references, and thus
will not be repeated here.  The CO data were converted to physical
units, namely solar masses per square parsec, using the same procedure as
Henry et al. (2003) who adopted a CO-to-H$_{2}$ conversion 
factor (from antenna temperature to mass) of $2 \cdot 10^{20}$ H$_{2}$
cm$^{-2}$ (K km s$^{-1}$)$^{-1}$ 
(Strong \& Mattox 1996).  Cloud sizes were determined by measuring the
major and minor axes of each cloud and calculating the
average. Fig.~\ref{fig:bima} shows the positions of the cluster complexes
on top of the {\it BIMA} intensity map.

\begin{figure}[h]
\begin{center}
\caption{{\it BIMA} CO intensity map, dark indicates regions of high
intensity.  North is up, and east is to the left.  The positions of
the cluster complexes are shown as
diamonds.  Note that most of the complexes fall on the outer edges of the spiral arm pattern.
The lack of spatial correspondence between the complexes and high CO
intensity regions indicates that the complexes have already destroyed
their parent GMCs.}  
\label{fig:bima}
\end{center}
\end{figure}

\section{Properties of the complexes}

\subsection{Sizes of the complexes}
\label{sizes}
\begin{figure}[h]
	\includegraphics[width=8cm]{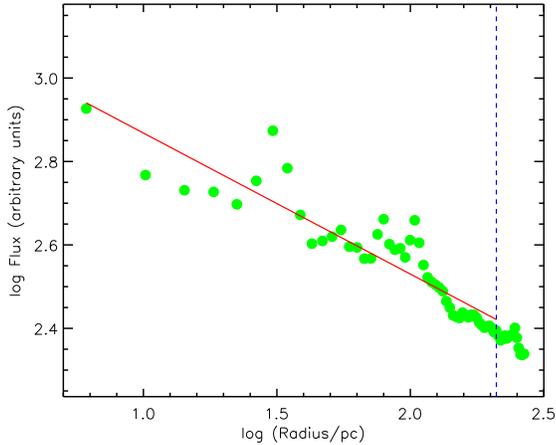}
         \caption{The flux in concentric circles of Complex
	G2.  The solid red line represents a profile where flux
	$\propto r^{0.34}$, while the vertical dashed line
	marks the adopted size of the complex.  The two bumps in
	the profile are bright sources within the complex.  All
	complexes studied here have similar power-law flux profiles
	(see text).} 
	%{\bf Right:}  The 
	%colours of the same 
	%complex as a function of distance from the centre.  The
	%vertical dashed line is the adopted radius of the complex.}
	       \label{fig:g-sb}
	       \end{figure}

The size of the complexes were measured on the {\it HST-WFPC2} images
and determined through the method used for 
star clusters, defined by Ma\'{i}z-Apell\'{a}niz (2001).  The method
defines the edge of the complex at the point where the colour, as a
function of radius, becomes constant.  To determine this point, we assumed
that the complexes are circular, and measured the surface brightness
in concentric rings.  This provides us with surface brightness and
colour profiles for each complex.  An example of the flux distribution
of one complex,
Complex~G2, is shown in Fig.~\ref{fig:g-sb}.  The vertical dashed line 
in the panel shows the radius adopted for this complex.  The
sizes of the complexes with U-band data (F336W) are shown in
Table~\ref{table:info}. 

It is important to note however, that the individual sources within a
complex do not show any colour dependence on their position.  The trend
seen in Fig.~\ref{fig:g-sb} is based on the {\it integrated} light within
each radius, which includes the background.  As one looks along the
radius away from the centre, the background light begins to dominate
untill the measured colour is equal to that of the background, which
we take as the radius of the complex.

\subsection{Intensity profile of the complexes}
\label{profile1}

In Fig.~\ref{fig:g-sb} we show the flux in
concentric cirlces in complex~G2.  We have over-plotted (solid
line)
a function of the form, Flux~$\propto r^{-0.34}$.  A power-law
of this form follows the data 
quite well (the two peaks at log radius of 1.5 and 2.0 in this
distribution are due to two 
clusters).  This corresponds to the projected density profile of a 
clump with $\rho(r)~\propto~r^{-\alpha}$, where $\alpha$ is 1.34 and
 $\rho$ is in units of \Msun pc$^{-3}$.  The flux profiles of the
other complexes in
our sample are also well fit by power-laws of this form, with the average
index, $<\alpha>~=~1.74~\pm~0.34$.  We will discuss the
physical implications of this in \S~\ref{profile2}.

\begin{table*}[h]%[!p]
{\scriptsize
%\begin{sideways}
\parbox[b]{9cm}{
\centering
\caption[]{Properties of the complexes in M~51 with F336W and H$\alpha$ imaging.}
\begin{tabular}{c c c c c c c}
\hline\hline
\noalign{\smallskip}
ID  & M$_V$$^{a}$ & M$_{V,cluster}$$^{b}$ & Mass$^{c}$  & Radius & Tidal Radius & SFR/area\\
   & (mag)  & (mag) &($10^{5} M_{\odot}$)& (pc) & (pc) &(\msun~yr$^{-1}$~kpc$^{-2}$)\\
\hline
\\
 
C2    &  -12.6 & -10.2 &1.4 & 160 &      37     &2.59 \\
D2   &	-12.4 & -9.8$^{d}$ &1.1 & 160 &      35    &0.07\\
E2    &	-11.2  & -9.0 & 0.4 & 85  &      27     &0.06 \\
F2    &	-11.1  &-9.0  &0.3 & 100 &      26    &0.06\\
G2    &	-13.3 & -10.1 &2.6 & 220 &      62     &0.07\\
B1    &	-12.0  & -10.4&0.7 & 125 &      36    &0.06\\

\noalign{\smallskip}
%\\

\noalign{\smallskip}
\hline
\end{tabular}
\begin{list}{}{}
\item[$^{\mathrm{a}}$] The magnitude of the complex within the defined
radius. Uncorrected for extinction.
\item[$^{\mathrm{b}}$] The magnitude of the brightest source within
each complex. Uncorrected for extinction.
\item[$^{\mathrm{c}}$] Total mass of the complex assuming an age of 7 Myr.
\item[$^{\mathrm{d}}$] Not shown in Fig.~\ref{ssc-colours} because it appears to
be a blend of sources, hence the colours are highly uncertain.

\end{list}
\label{table:info}
}
%\end{sideways}
}
\end{table*}

\subsection{Ages of the complexes}
\label{sec:ages}

\subsubsection{Ages determined by H$\alpha$ measurements}
\label{ageew}

The presence of H$\alpha$ emission within the complexes indicate that
they are quite young, and as such suffer from the degeneracy between age and
extinction (e.g. Bastian et al. 2005b).  We therefore first determine the
ages of the complexes (with F336W data) using the H$\alpha$ emission line
width, which is independent of extinction.

To estimate the equivalent widths of H$\alpha$ we use the F656N narrow band
filter. The F675W filter was used to estimate the continuum contribution.
Photometry was performed using apertures set to be as large
as the estimated radius of each the complex. The monochromatic flux in
the F656N and F675W band can then be
found as follows:
$F_\lambda = SUM_\lambda * PHOTFLAM_\lambda / EXPTIME_\lambda$, where
$\lambda$ refers to the central wavelength of the filter, 
$SUM_{\lambda}$ is the sum of the counts within the aperture
(i.e. without background subtraction) and $PHOTFLAM_\lambda$ and
$EXPTIME_\lambda$ were taken from the image headers.

The equivalent width of H$\alpha$ can now be found by
                                                                               
\begin{equation}
EW(H\alpha) = 28.33 * \frac{F_{H\alpha} - F_R}{F_R}
\end{equation}
where 28.33 is the rectangular width of the H$\alpha$ filter in \AA.
                                                                               
The errors in EW(H$\alpha$) were estimated by
calculating the EW values with apertures of plus and minus 5 pixels, which is
approximately the uncertainty in the radius estimates. 

\begin{figure}[h]
\begin{center}
\includegraphics[width=8cm]{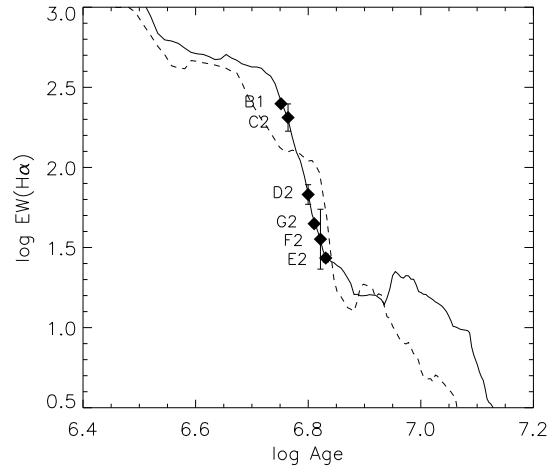}
\end{center}
\caption{Determining the ages of the complexes from the observed
        equivalent width of H$\alpha$ (EW(H$\alpha$)) combined with the
        Starburst99 
        models (Salpeter IMF).  The solid line is for solar
        metallicity while the dashed line is for twice solar
        metallicity.  The EW of each
        complex was put on the line, and the corresponding age is read
        off.  The errors were calculated by varying the apertures
        used for the photometry by 5 pixels (plus and minus) around
        the measured radius.  Note that the derived ages are
        largely independent of the assumed metallicity.}  
\label{ew}
\end{figure}

We can then compare the derived EW for each complex, with {\it
Starburst99} simple stellar population (SSP) models for solar
metallicity and Salpeter IMF
(Leitherer et al.~1999), in order to derive the age of each complex.
The results are shown in Fig.~\ref{ew}.  We note that the assumed
metallicity ($Z_{\odot}$ or $2Z_{\odot}$) does not significantly
affect the derived ages.  We see that the complexes are
indeed very young, with ages between 5 and 8 Myr.  In
\S~\ref{ageclusters} we will show that the measured cluster colours
are well matched by cluster models of solar and twice solar
metallicity.  

This method assumes that each complex formed in an
instantaneous burst, hence it puts a lower limit on the age
of the complexes, as any additional formation of massive stars will
tend to increase the equivalent width of H$\alpha$.  This means that a
combination of an older burst plus new star formation can mimic the
observed strength of the equivalent width.

Along similar lines, we can also look for the presence of young O and
B stars at the location of the complexes.  For this we compare the
location of the complexes (found in the optical) to far-{\it UV}
images taken by the {\it GALEX} observatory.  Fig.~\ref{fig:galex}
shows the composite {\it HST-Hubble Heritage} image (top) as well as the
composite far-{\it UV} (1530\AA,~blue) and near-{\it UV} (2310\AA,~red)
{\it GALEX} image (bottom, taken from Bianchi et al. 2005) of the
north western inner spiral arm. 
We find strong far-{\it UV} flux at the location of most of the
complexes which is a further indication of the presence of young stars
within the
complexes.  Also note the semi-regular spacing of star-forming regions
along the spiral arm, the so-called ``beads on a string'' morphology.
It appears that the largest isolated star-forming regions are
outside the {\it HST} field of view, being located in the south-west
and north-east of the galaxy.  These complexes will be the subject of
a future study which will exploit the full coverage of M51 with the
{\it HST-ACS} mosaic.

A comparison of Fig.~\ref{fig:bima} and Fig~\ref{fig:galex} reveals the
different 
stages of evolution of gas/star complexes in the NW arm, from A2 to
G2.  The H~II complexes to the south-east of Complex~C2 are seen in the
{\it UV} but are not present in the CO data.  Contrarily, the H~II complex
to the west of complex~D2 is unseen in the {\it UV} but is bright
in CO.  Presumably, we are witnessing complexes in different stages of
their evolution from CO complexes, to large H~II regions, and finally
to star/cluster complexes.

\begin{figure}[h]
\begin{center}
\caption{{\bf Top:}{\it Hubble Heritage} image of the north western
inner spiral arm of M~51.  {\bf Bottom:} {\it GALEX} composite
far-{\it UV} (blue) and near-{\it UV} (red)
image of the same region.  Bright regions indicate strong {\it UV}
flux, indicating the presence of hot O and B stars. Note the
semi-regular spacing of the complexes along the spiral arm in both
images.  North is up and east is to the left in this image.}  
\label{fig:galex}
\end{center}
\end{figure}

Finally, we note that complex~G2 is located at a break of a strong
CO/dust lane and at the beginning of a large spur.  Both of these
observations may be hints into its formation.  The position of G2 near
the outer edge of the stellar arm might be explained by its age;
however, we do not find any strong correlation between the distance
from the arm edge and the age for the other complexes.

\subsubsection{Ages from the individual clusters}
\label{ageclusters}
\begin{figure*}[htb]
\includegraphics[width=18cm]{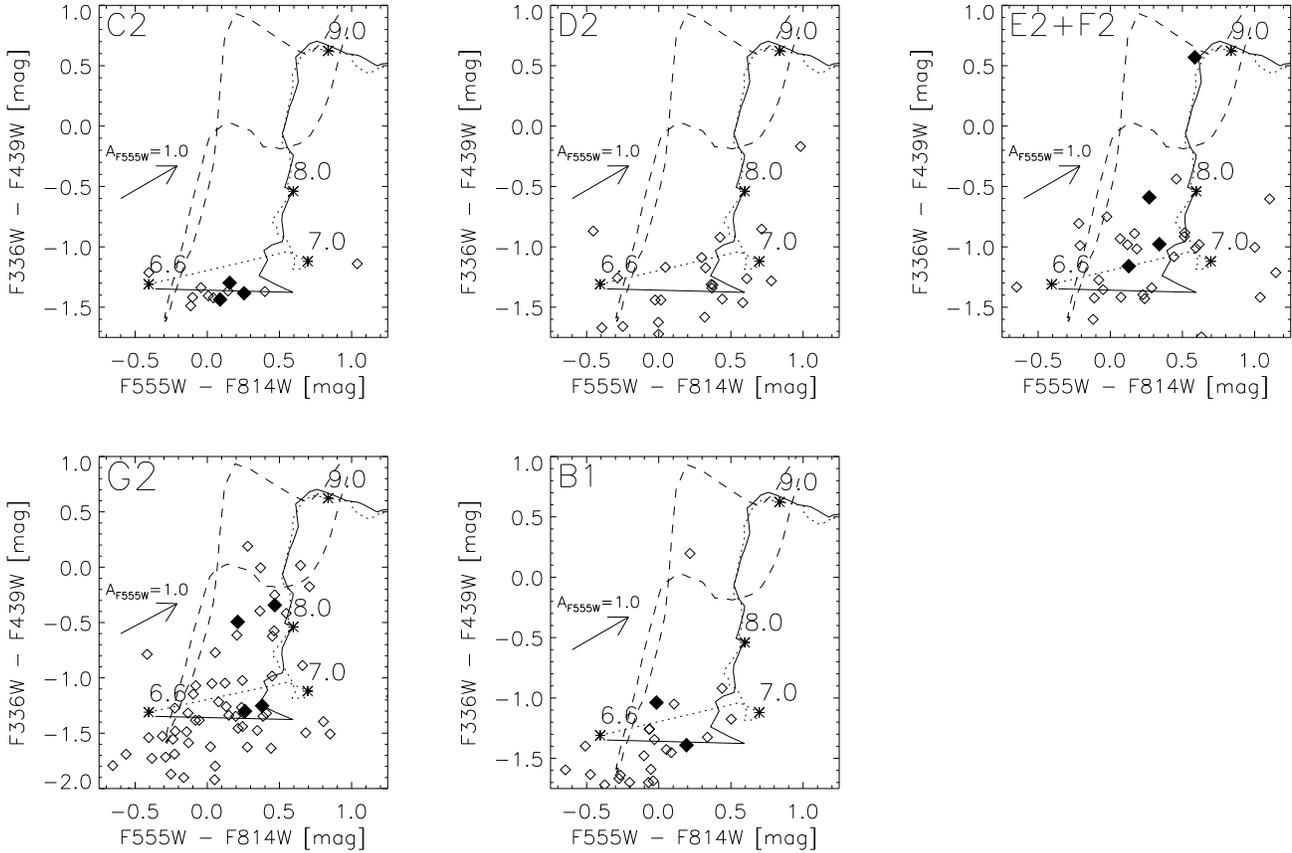}
\caption{Colour-Colour diagrams for sources in the six cluster complexes
        with F336W observations.  The names of the complexes are given
        in each panel.  Filled points are for sources with
        $M_{V} < -8.6$ (uncorrected for extinction), while open points
        are fainter than this (somewhat arbitrary) limit.  The
        solid line represents the {\it GALEV}~SSP models for twice
        solar metallicity while the dotted line represents the same
        models but for solar metallicity.  The asterisks mark age
        points on the evolutionary curves, labelled as the logarithm
        of the age in years.  The long dashed line is a stellar
        isochrone for a solar metallicity, 10 Myr old, population
        (however the isochrone for twice solar metallicity is extremely
        similar in this region of colour space).  The arrow in each
        panel represents the extinction correction of $A_{\rm
        F555W}$=1.0. }  
\label{ssc-colours}
\end{figure*}

We can also determine the ages of the complexes by looking at the ages
of the individual star clusters within the complexes.
Fig.~\ref{ssc-colours} shows (F336W - F439W) vs. (F555W - F814W) for
the sources within the six complexes for which we have F336W
observations, which are essential for the age dating of young star
clusters (e.g. Anders et al. 2004).  We also show the solar (dashed-dotted
line) and twice 
solar metallicity (solid line) {\it GALEV} SSP model tracks for
a Salpeter IMF.  The filled data points are clusters
with $M_{\rm F555W} < -8.6$, which are highly likely to be star clusters.
The open points are the fainter sources within each complex, which may
be faint clusters or individual bright stars.  Assuming, for the
moment, that the 
sources within the complex are star 
clusters, we see that the 
majority of the clusters have ages of $\sim4 - 10$ Myr.  The 
exact age is difficult to determine due to age, extinction, and
metallicity degeneracies for young clusters (e.g. Bastian et
al. 2005b).  However, we conclude that the ages
derived from the H$\alpha$ equivalent widths of the complexes are
consistent with the ages derived from the colours of the individual clusters.

 In Fig.~\ref{ssc-colours} we see that complexes~C2, D2, and B1 have a
 relatively small amount of scatter of their 
points around the extinction vector, implying similar ages between all
the clusters.  The other complexes (E2, F2, and G2),
 however, contain some sources that appear 
much older ($>$~100 Myr) in the colour-colour diagram.  These sources
 may in fact be young clusters which are 
heavily extincted or individual massive stars. In particular the sources in
G2, show a  significant amount of scatter perpendicular to
 the extinction vector. Taken at face value this
suggests that this complex has been forming clusters for at least the
last $\sim100$ Myr.  

\begin{figure}[htb]
\hspace{-0.25cm}\includegraphics[width=9cm]{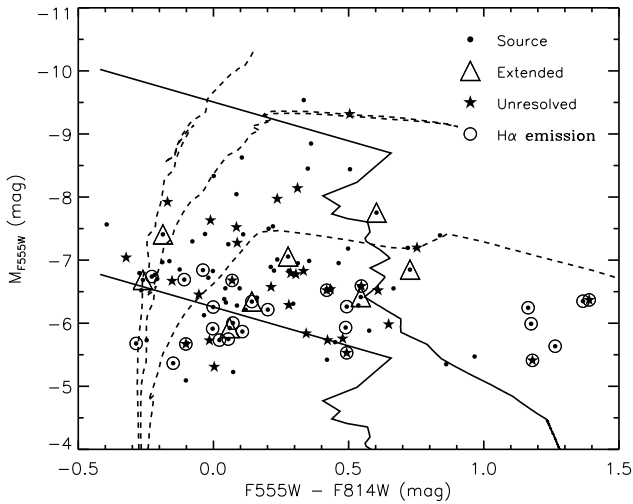}
\caption{Colour-magnitude diagram of the sources within complex~G2
(from the {\it ACS} images).
        The dotted lines are solar metallicity stellar isochrones for
        3, 5, and 10 
        Myr from left to right (the twice solar metallicity isochrones
        occupy the same region in this figure). The solid lines are {\it GALEV}~SSP
        (cluster) model tracks
        (Salpeter IMF and twice solar metallicity) for a $10^{4} M_{\odot}$
        (upper) and $500 M_{\odot}$ (lower) cluster.  The open
        triangles denote sources which appear extended, the
        stars are sources which are not resolved ($\leq~1$~pc), open
        circles are sources which have strong H$\alpha$ emission
        associated with them, and filled dots are detected sources
        which have no size information (due to possible crowding or
        low S/N).}   
\label{g-cmd}
\end{figure}

Some of the sources which appear quite old in complexes~E2, F2, and
G2, however, may be individual 
bright stars.  In order to test this, in Fig.~\ref{g-cmd} we plot
the absolute magnitude of each source within complex~G2 (uncorrected
for extinction) vs. the colour (F555W - F814W, roughly V-I).  The solid
lines are {\it GALEV} SSP model tracks for a $500$\msun (lower) and
$10^{4}$\msun 
(upper) cluster.  The dashed lines are stellar
isochrones (with solar metallicity) of  3, 5,
and 10 Myr (from left to right respectively, Lejeune \& Schaerer
2001). Open triangles show sources which appear extended, filled
stars are sources which are not resolved ($\leq 1$~pc), open circles
denote sources with strong H$\alpha$ associated with them, and filled
points are sources with no size information (either because of
possible crowding or low signal-to-noise).   We see that
many of the objects in complex~G2 are 
consistent with both the stellar isochrones and the cluster model
tracks. We note however that there is a significant grouping of
sources that would be consistent with star clusters of a few~$\times
10^{2}$\msun to $\sim~10^{4}$\msun and ages between 4 and
10 Myr (between $-0.3$ and $0.5$ in F555W-F814W).

Stochastic sampling of the underlying stellar
IMF can cause significant deviations from standard SSP model colours
(e.g. Dolphin \& Kennicutt 2002), therefore it is not completely
unexpected that sources do not lie directly on the SSP model tracks.
We also note that due to the large 
crowding in these complexes, accurate photometry is difficult and will
tend to increase the scatter in diagrams such as Fig.~\ref{g-cmd}.
Without higher resolution imaging or spectroscopy of individual 
sources, the degeneracy between stars and star clusters is extremely
difficult to break. 
 
Based on the brightness criterion of M$_V < -8.6$, we conclude that
there are clusters within the complexes and that their ages are between 4
and $\sim$10 Myr, although complexes~G2, E2, and F2 seem to also
contain older clusters.

It is interesting to note the existence of spatially
resolved massive star clusters within the complexes, the most massive
of which is located at the centre of complex~A1.  Using the available
{\it HST-ACS} images, we find that this cluster has a magnitude of {\it
m(F555W)}$_{\rm ACS}$=18.3, and a colour {\it m(F435W) - m(F555W) =
0.32}.  This, together with the strong H$\alpha$ emission, is
consistent with the cluster being quite young and extincted.  Using
the same model and distance assumptions used throughout this work, and
an assumed age of the cluster of 7~Myr, we find that this cluster has a mass
of $\sim 10^5 M_{\odot}$, which is a lower limit as we have not
corrected for extinction.  Using the {\it ISHAPE} routine of Larsen
(1999), we find this cluster to be well resolved, although significantly
flattened.  Using a Moffat profile with index 1.5, we find the FWHM
along the major axis of the cluster to be 2.6$\pm$0.5 pc, and a major
to minor axis ratio of 0.6.  In a future work, we will present an in
depth analysis of the cluster and stellar populations within each
complex.  Another interesting feature of this complex, are two dust
arcs which appear to have the same centre, which is
displaced from the young central massive star cluster (shown in the
bottom panel of
Fig.~\ref{g-rgb}).  The origin of these arcs remains unclear.

%{\bf As the bottom panel of Fig.~\ref{g-rgb} demonstrates, there is no
%strong H$\alpha$ emission from the central large star cluster in
%Complex A1 itself, although the edges surrounding the complex bright
%in H$\alpha$ emission.  The double HII supershells seem to be due to
%the gas expelled from the complex by the pressure induced by the
%central star cluster.  The blue colour (i.e. low extinction and young
%sources) of the inner regions of the complex is further evidence that
%the gas has been efficiently removed from the complex......fix!!!}

\subsection{Masses of cluster complexes}
The masses of the complexes were estimated by combining the complex
luminosities and the age dependent mass-to-light ($M/L$) ratios of the {\it
GALEV} SSP models.  We assign all complexes the same age, $\sim7$ Myr,
and interpolate the SSP models between 4 and 8 Myr to determine
$M/L$ for this age.  We have used the F555W magnitudes to derive the
masses of the complexes, but we note that the other broad-band
magnitudes give similar results.  The masses of these complexes range
from $0.3 \cdot 10^{5}$\Msun~(F2) to $3.0 \cdot
10^{5}$\Msun~(G2) and 
are given in Table~\ref{table:info}.  The errors on the determined masses (as
seen in Fig.~\ref{m-v-r-complex}) are based on errors in the age
estimate, from 4 to 12 Myr.

\subsection{Star formation rates within the complexes}
\label{sec:sfr}

Using the continuum subtracted \halpha  images, we can estimate the
star formation rates, $\Sigma_{\rm SFR}$,  within the complexes.  For
this, we adopt the 
conversion factor of \halpha luminosity to star formation rate of
Kennicutt (1998a), namely
$$\Sigma_{\rm SFR}(M_{\odot}~{\rm yr}^{-1}) = 7.9 
\cdot 10^{-42} L(H\alpha)({\rm ergs~s}^{-1}).$$
The values derived are shown in Table~\ref{table:info}.  The star
formation rates per unit area of the complexes are comparable to the
definition of a starburst galaxy (0.1 \msun~yr$^{-1}$~kpc$^{-2}$, Kennicutt
et al. 2004), justifying complex~C2 as a  {\it localized starburst}
(terminology from Efremov 2004).  Other complexes from Table~\ref{table:info}
have lower SFRs.  

Star cluster complexes in the Antennae galaxies have area normalised
star formation rates which are approximately 10 times higher
than the ones studied here, despite having similar sizes (Bastian et
al. 2005b).  This difference can be understood by comparing the GMCs
properties in the two galaxies.  The resolved GMCs in the Antennae (Wilson et
al.~2003) have roughly 10 times the mass, and hence 10 times the
density, as those in M51 for a given spatial size. Therefore, the
differences in the SFR/area can be well explained through the Schmidt
law (Schmidt 1959) where the rate of star formation is proportional to the gas
density to the power $N$.  Kennicutt (1998b) has shown that
$N$~$\sim~1.4~\pm~0.15$ for normal disk galaxies.

\section{Comparison between the properties of cluster complexes and GMCs}
\label{sec:comparison}

\subsection{The density profiles}
\label{profile2}

In \S~\ref{profile1} we found that the surface brightness of complex~G2
is well fit by a power-law of the form
Flux~$\propto~r^{-0.34}$.   Assuming sphericity, this
corresponds to a three dimensional density profile of
$\rho~\propto~r^{-\alpha}$, with $\alpha=1.34$.  The other complexes
were also shown to be well fit with profiles of this type, with
$<\alpha>~=~1.74\pm0.34$.  This power-law profile is similar to that
observed for GMCs ($\alpha=1-2$, Ashman \& Zepf 2001)\footnote{Recent
studies have shown that Galactic GMCs have $\alpha=1.7 \pm 0.2$, Joao
Alves priv. comm.} and much shallower than King profiles which fit
old globular clusters quite well.  The similarity between the profiles
of GMCs and the cluster complexes studied here suggests that the
amount of luminous material formed is proportional to the gas density
at that point within the GMC.  We shall return to this point in
\S~\ref{sec:formation}.  Throughout the present work we assume that
the complexes are spherical, which in the z-axis would make the
complexes larger than the thickness of typical galactic disks.  This
is also true for the GMCs.  If both the GMCs and the complexes are truncated in
the vertical direction (i.e. flattened) our main conclusions would
still remain valid.

\subsection{Size and radius relation of GMCs and cluster complexes}
\label{sec:mass-radius}

By combining the study of the cluster complexes and GMCs in M51 we can
gain insight into the hierarchical nature of structure formation within
spiral galaxies.  Solomon et al. (1987) showed that there is a clear
relationship between size and mass for GMCs in the Galaxy, namely
$M_{\rm Cloud} \propto R_{\rm Cloud}^2$, a consequence of
virial equilibrium.  This relation has also been found for GMCs
outside the Galaxy (see summary in Ashman \& Zepf 2001).  

\begin{figure}
\includegraphics[width=8cm]{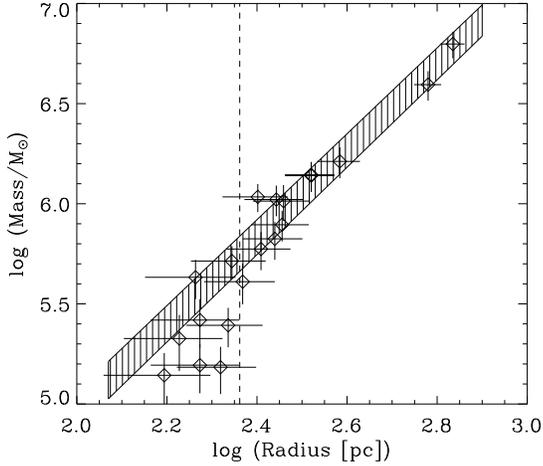}

\caption{The mass vs. radius relation for GMCs in M51.  The vertical
        dashed line indicates the spatial resolution of the {\it BIMA}
        survey.  The filled area is a power-law fit to the data above
        the resolution limit, with index 2.16, and the corresponding
        $1\sigma$ error bars on the index ($\pm 0.20$) and the zero point.}
\label{m-v-r-cloud}
\end{figure}

However, as shown here, a single massive GMC does not produce a single star
cluster, but a complex of star clusters.  Therefore it is interesting
to see how these complexes fit into the hierarchy of star formation.
We begin by searching for a relation between mass and radius for GMCs
in M51. Fig.~\ref{m-v-r-cloud} shows the radius vs. mass for the
detected GMCs within M51.  The dashed vertical line shows the spatial
resolution of the {\it BIMA} survey.  We have fit a function of the
form log~$M$~=~$\kappa~\cdot$~log~$R$~+~$c$
(i.e.~$M$~$\propto$~$R$$^{\kappa}$).  We find  
that $\kappa = 2.16 \pm 0.20$.  The filled area corresponds to the best
fit to the data above the resolution limit including the $1\sigma$
error (this area also includes the error on the zero point of the function).

Fig.~\ref{m-v-r-complex} shows the result of the same analysis applied
to the cluster complexes.  Here we find that $\kappa = 2.33 \pm 0.19$.
The filled area with the horizontal hash marks represents the best
fit plus the corresponding $1\sigma$ errors.  The filled area with the
vertical hash marks is the fit (plus errors) of the GMCs extrapolated into
the size range of the complexes.  We see that the relations for the
GMCs and the cluster complexes have almost the same slope, but are
offset by $\sim 0.3$ dex in the vertical (mass) direction.  This
offset is related to the star formation efficiency which will be
discussed in \S~\ref{sec:efficiency}. 

{\it We conclude that the cluster complexes follow a similar
mass-radius relation as GMCs.  On the contrary, young star clusters do
not follow this mass-radius relation (e.g. Bastian et al. 2005a).
This implies that the 
mass-radius relation must be broken below the scale of the complexes.}
If binding energy is 
the factor that determines the star formation efficiency (assuming
that the star formation efficiency causes the lack of a
mass-radius relation in young star clusters) then it is not the
binding energy of the cloud which is important but the binding energy
of the clumps within the cloud which is the dominant factor.  

Additionally we note that the mass-radius relation holds for clumps
within GMCs, down to the resolution limit ($\sim0.5$ pc and $\sim
13$~M$_{\odot}$ in the Rosette molecular cloud (Williams et
al. 1995).  This suggests that the mass-radius relation is broken
during (or after) the star formation process.

\subsection{Star formation efficiency within the complexes}
\label{sec:efficiency}

We have shown that the GMCs and the star cluster complexes in M51
share a very similar mass vs. radius relation.  This offers a unique
opportunity to investigate the star formation efficiency, $\epsilon =
M_{\rm complex}/M_{\rm GMC}$, for each complex.  Although we cannot
measure the mass of the GMC that formed the present complexes, if we
assume that the parent GMCs had the same size as the complexes, we can
estimate the progenitor mass.  Thus it remains to be shown that the
radius of the clouds do not change significantly during the formation
of the complexes.

We can estimate the timescale on which the GMCs will shrink by
calculating their free fall time. The free fall timescale is $t_{\rm
FF} \approx
(\rho \cdot {\rm G})^{-\frac{1}{2}}$.  Using the relation between the
mass and radius of the clouds derived above, we see that the free fall
timescale of GMCs, as a whole, is between 80 and 200 Myr.  The ages
(e.g. the upper 
limit of the formation timescale) of the complexes are between 5 and
10 Myr, much shorter than the collapse time of the GMCs.  This means
that the complexes formed from the GMCs on a timescale much shorter
than the collapse time of the GMC.  Thus the size of the complexes
should be about the same as that of the GMC from which they formed.

\begin{figure}
\hspace{-0.3cm}\includegraphics[width=8.5cm]{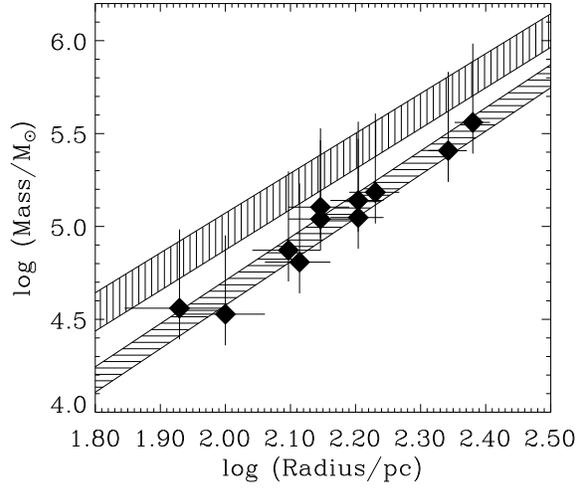}

\caption{The mass vs. radius relation for cluster complexes in M51.
        The solid line is a power-law fit to the data with index,
        2.3.  The filled area with vertical hashes is the fit to the GMCs from
        Fig~\ref{m-v-r-cloud}, extended into the size regime of the
        complexes.  The filled area with the horizontal hash marks is
        the power-law fit to the complexes, with index 2.33, and the
        corresponding $1\sigma$ error bars on the index ($\pm 0.19$)
        and the zero point.}
\label{m-v-r-complex}
\end{figure}

In Fig.~\ref{m-v-r-complex} we show the mass-radius relation for the
complexes (the data points along with the best fitting power-law
relation with the associated $1\sigma$ errors) and over-plot the
relation for the GMCs, extrapolated to the size scale of the
complexes.  The vertical offset between the best fit line of the complexes
and the best fit line of the 
the GMCs is the star formation efficiency, $\epsilon$.  In the case of
M51, we measure $\epsilon$ of the complexes to be
50 $\pm$ 20 \%.  The convergence of the two relations at higher masses
suggests that there is a trend for $\epsilon$ to increase for
larger complexes, but the trend is well within the observational
errors.  

An additional caveat to this approach is that we have assumed the
mean Galactic CO-to-H$_2$ conversion factor.  However, studies of M51
have shown that the conversion factor in M51 is roughly half that of
the Galaxy (Boselli, Lequeux, \& Gavazzi 2002).   Using this value
would make GMCs half as massive as that estimated here, and therefore
the star formation efficiency would be close to 100\%.  A final caveat
to this estimate is the assumed stellar initial mass function for
the complexes.  We have assumed a Salpeter IMF from 0.1 to 50
$M_{\odot}$, however if we would assume a Kroupa-type stellar IMF, the
masses of the complexes would be lowered by a factor of $\sim2$,
giving $\epsilon = \sim25$\%.  Therefore we are left to conclude that
while this method to estimate the star formation efficiency is
intriguing, it is limited at the present time by the assumptions that
are required.

 However, it should be noted that the assumed CO-to-H$_2$ factor
will not effect the index of the mass-radius relation of the GMCs
presented here. 
The GMCs in M51 used in this study are all located within the inner
5~kpc of the galaxy, and therefore we do not expect much cloud to
cloud variation in the conversion factor, as the metallicity is not
expected to change significantly over this area.

\section{Discussion}
\label{discussion}

\subsection{Formation of the complexes}
\label{sec:formation}
We can compare the properties of the complexes in M51 with those
of complexes and young/old star clusters in the Antennae
galaxies which have been measured by Whitmore et
al. (1999). Knot~S\footnote{The large amount
of substructure within this knot, particularly in the F555W
($\approx$V) and H$\alpha$ bands, suggests that this knot is a
cluster complex, and not just a single large star cluster.} in
the Antennae  galaxies has an extremely similar 
power-law profile (Flux~$\propto~r^{-1/2}$) over $\sim 300$ pc from
the centre as complex~G2.
Additionally, the very young ($<$ 10 Myr) massive star 
cluster \#430 in the Antennae follows the same power-law density
distribution.  However, the older cluster \#225 ($\sim500$ Myr old)
does not follow a clear power-law relation but instead
has a sharp cut-off at $\sim50$ pc.  Presumably cluster \#225 was
formed with a power-law density profile which has been eroded due to
dynamical evolution.  Additional support for this scenario comes from
star clusters in the LMC, where young star clusters show a power-law
density profile (with no distinct cut-off radius)(Elson, Fall, \&
Freeman 1987) while older clusters have King-type profiles with a
distinct tidal truncation. 

The similarity between the profiles of GMCs and the
complexes studied here suggests that the amount of luminous material
formed at a certain radius within a GMCs is proportional to the
density of the cloud at that radius.  This naturally explains why the
complexes share the same mass-radius relation as GMCs (see
\S~\ref{sec:mass-radius}), 
as any relation inherent to the progenitor cloud will be imprinted onto
the complexes as they form.  Additionally, the similarity between the
projected profiles of the complexes, and young clusters implies a
common formation mechanism (i.e. star formation proportional to the
gas density). 

However, young cluster systems, such as those of NGC 3256 (Zepf et
al.~1999), M51 (Bastian et al. 2005a), and various spiral galaxies
(Larsen 2004) have a weak relation between their mass and radius (with
a large scatter).  The similarity between the density profiles of the
complexes and young star clusters makes the lack of a mass-radius relation
in star clusters all the more surprising.  This is because we would also expect
them to bear the imprint from the cloud of which they formed.  

Ashman \& Zepf (2001) have suggested
that a star formation efficiency which depends on the binding energy
(predicted by Elmegreen \& Efremov 1997)
of the progenitor GMC could destroy such a relation during the
formation of young clusters.  But this theory does not explain the
shallower size distribution of star clusters relative to GMCs.

Another possible explanation for this lack of a mass-radius relation in young
clusters is that dynamical encounters between young clusters (and gas
clouds) add energy into the forming 
clusters, thereby increasing their radii. Some support for this scenario
is given by differences in the size distributions of clusters and GMCs. 
GMCs (e.g. Elmegreen \& Falgarone 1996) and cloud clumps
(e.g. Williams et al. 1995) follow the size 
distributions of $N(r)dr \propto r^{-\eta}dr$ where $\eta~\approx~3.2$.
Star clusters (both young and old), however, follow a shallower
relation, namely a power-law with $\eta \approx 2.2$ (Bastian et
al. 2005a).  If the proto-clusters follow the same mass-radius relation
as GMCs and clumps, then their density will be $\rho \propto M/R^{3}
\propto R^{-1}$, showing that the larger proto-clusters are less
dense.  Due to their lower density, we expect larger proto-clusters to
be more affected by encounters, i.e. making large clusters even
larger.  This would tend to make the size distribution shallower, as observed.

\subsection{Truncation of the size of the complexes}
The complexes most closely associated with spiral
arms (e.g. excluding complexes~C1 and G2) have diameters similar to
the minor axes of GMCs within the arms.  We have shown that the size of
a complex does not change significantly from the progenitor GMC.  This
suggests that 
the maximum size (and hence mass due to the 
mass-radius relation) of the complexes is determined by the size of
the GMCs within the galaxy.  Due to sheer effects, caused by the flat
rotation curve of the disk, GMCs have a maximum size (hence mass).  This
in turn will determine the maximum sized complex which can form.  If
the star clusters within the complex are formed with an initial mass
function, then the most massive cluster formed will be
dependent on the size of the parent complex.  There does seem to be a
relation between the most massive GMC and the most massive star
cluster within a galaxy (e.g. Wilson et al.~2003). 

However, there are two exceptions where the size of a complex is
larger than the typical semi-minor axis of a GMC.  These two complexes
(A1 and G2) do not seem to be directly related with a spiral
arm, in fact A1 is on the inside of the spiral arm.  G2 is located at
the starting point of a large spur, which is a sign of instabilities
in the arm.  Instabilities, such as large spurs, may
allow larger GMCs to exist for short periods, and hence explain the
existence of these larger than expected complexes.

\subsection{The evolution of the cluster complexes}

As shown in \S~3.2, the complexes studied here are all quite
young.  However, if the complexes are a long lived phenomenon, then we
would expect to see complexes throughout the disk and not just
associated with the spiral arms.  Elmegreen (1994) predicted that the
remnants left over from 'superclouds' within the spiral arms,
(i.e. what we call complexes) should be observable as loosely bound
stellar complexes in the inter-arm region of the galaxy.  Do such
inter-arm complexes exist?  The 
present data set is not adequate to conclusively answer this
question.  As shown in Bastian et al. (2005a) the detection limit
imposed by the data severely limits which clusters (e.g. how bright
and hence how massive) we can observe.  At an age of $\sim100$ Myr, we
can only detect clusters (with no extinction) which have masses above
$10^4$\Msun.  Therefore, we would miss smaller clusters belonging to
the same complex, if they exist.  We have found 3 candidate complexes
with ages between 40 and 80 Myr as determined by the colours of the
detected clusters.  Deeper images are necessary to resolve this question.

In order to estimate whether or not the complexes studied here are
gravitationally stable, we can compare their present radii with their
estimated tidal radii.  In a rotating disk, this can be estimated by
$$r_{t} = (\frac{G M_{\rm complex}}{2 \cdot V_{G}^2})^{1/3} {
R_{\rm G}^{2/3}} $$
where $M_{\rm complex}$, $V_{G}$, and $R_{G}$ are the mass of the complex, the
circular velocity of the galaxy at that point, and the distance to the
galactic centre.  If we assume a disk
rotation velocity of 200 km/s (e.g. Rand 1993 for galactocentric
distances greater than 1~kpc) and distances to the
galactic centre between 2 and 4 kpc, then we see that complexes
with masses between $10^{4} M_{\odot}$ and $10^{5.5} M_{\odot}$ have
tidal radii of 50 to 100 pc.  In Table~\ref{table:info} we compare the
measured sizes of the complexes with their derived tidal radii.  We
see that the tidal radii are much smaller than the measured radii,
indicating that the outer material of the complexes are not bound to
the complex.  Because of the steep density profile of the complexes
(see Fig.~\ref{fig:g-sb}) the inner region may be bound and survive
for extended periods.  If this is the case, then the inner clusters
are expected to merge within a few $\times 10^{7}$ years (e.g. Kroupa
1998; Fellhauer \& Kroupa 2002), forming a single massive star
cluster like object.

Elmegreen, Efremov, and Larsen (2000) and Larsen et al. (2002) have
speculated that the most massive cluster within 
the giant stellar complex in NGC~6946 may have formed by the merging
of other members of the complex.  This cluster is located in the
centre of the complex and its measured age (through photometry and
spectroscopy) appears to be the average of the surrounding clusters,
which is exactly as the merger theory predicts.  We note that the
centres of many of the complexes presented here, show the presence of
massive and elongated objects, in particular complexes~A1, E2, F2, and
G2. We take the presence of such objects as evidence for merging in
the centre of the complexes.  Higher resolution imaging should offer
more definitive evidence of on-going merging in the centres of these complexes.

\section{Summary and conclusions}
\label{conclusions}
\begin{enumerate}

\item {\bf Cluster complexes in M51:}  From archival
{\it HST} imaging we identified 9 cluster complexes in the inner
$\sim5$ kpc of M51.  We derived their ages by comparing their
equivalent widths of H$\alpha$ to that of {\it Starburst99} SSP
models.  We checked the validity of these age measurements by comparing
the colours of the star clusters within each complex to the {\it
GALEV} SSP models.  We find reasonable agreement between the two methods.

\item {\bf GMCs in M51:} We have also measured the sizes and masses of
a sample of giant molecular clouds in M51 from existing CO data.  This
was done in order to enable a comparison between the complexes and
the gas content of the galaxy.

\item {\bf The star formation rate within the complexes:} We have
measured the star formation rates within the complexes from 
the (continuum subtracted) H$\alpha$ flux using the relation between
the H$\alpha$ flux and the star formation rate given by Kennicutt
(1998a).  The area normalised star 
formation rates are as high as starburst galaxies, thereby justifying
their designation as {\it localized starbursts}, a label first coined by
Efremov (2004).  We found that the star
formation rates (for a given size) within the complexes studied here
are about a factor of 10 lower than complexes (of the same size) in
the merging Antennae galaxies.  We trace this difference to the 
molecular cloud populations within the two galaxies, as the GMCs in
the Antennae galaxies are roughly 10 times more dense than the GMCs in
M51.  Thus, if star formation is proportional to gas density, as is
the case in the commonly used Schmidt law of star formation, then the
difference in the star formation rates between the two galaxies is
readily understandable.

\item {\bf The surface density distribution:} The cluster complexes
follow the same surface density distribution as that 
of GMCs and young star clusters, namely a power law with exponent
$-0.74\pm0.34$, corresponding to a spatial density
$\rho~\propto~r^{-1.74}$. This provides a natural explanation for the
similarity
between the observed mass-radius distribution of GMCs and cluster
complexes.   Presumably older clusters follow a steeper density
distribution with a distinct tidal truncation due to dynamical evolution.

\item {\bf The mass-radius relation:} The complexes in M51 follow the same
mass-radius relation observed for GMCs. We conclude that this
similarity is due to the imprint of the progenitor GMC onto the
complex forming within it.  The shared mass-radius relation is a
natural consequence of the shared density distributions.  Contrary to
the complexes however, young star clusters do not share the same mass-radius
relation (they 
follow a much weaker relation with a large scatter).  We suggest that
this may be due to interactions between the young star clusters and gas
clouds.  This scenario is supported by the observation that the size
distribution is significantly shallower for star clusters (young and
old) than for GMCs (Bastian et al. 2005a).

\item {\bf Star formation efficiency:} We have argued that the size
of the progenitor GMC and that of the each complex are the same, due
to the short formation timescale relative to the free-fall timescale
of the GMC.  We estimate the star formation efficiency within 
each complex, by extending the observed mass-radius relation of GMCs
to that of the scale of the complexes.  In this way, we estimate that
the star formation efficiency in the complexes is 50\%.  However, due
to the required assumptions (CO-to-H$_2$ conversion factor and stellar
IMF of the complex) the errors using this method are quite
substantial, and hence this method is not viable at the present time.

\item {\bf Evolution of the complexes:} The complexes studied here are
quite similar to those modelled by 
Fellhauer \& Kroupa (2002, 2005).  Their
models show that a significant amount of merging of the individual
clusters  is expected to happen
within the complexes, which may lead to the formation of a much larger
star cluster than would be expected from statistical sampling of a
cluster initial mass function.  This may explain why clusters with
masses up to $5 \times 10^{5} M_{\odot}$ are found in M51 (Bastian et
al. 2005a), which is about the mass of the largest complex analysed
here.  Many of the complexes studied here show evidence of merging in
their centres.

\end{enumerate}

\begin{acknowledgements}
This research has benefited from the NASA Extragalactic Database. We
thank S${\o}$ren Larsen and Pavel Kroupa for
useful and stimulating conversations.  We would like to thank
Alice Quillen for help in analysing the {\it BIMA} data.
Yu.N.~E. acknowledges support from grants RFFI 03-02-16288 and NSh
389-2003-2.   

\end{acknowledgements}

\bibliographystyle{alpha}
\bibliography{../bib/astroph.bib,../bib/phd.bib,../bib/mark.bib}

\end{document}